\documentclass[a4paper,fleqn]{cas-dc}

\usepackage{amssymb}
\usepackage{lipsum}
\usepackage{url}
\usepackage[authoryear]{natbib}

\newcommand{\msun}{M_\odot}

\begin{document}
\let\WriteBookmarks\relax
\def\floatpagepagefraction{1}
\def\textpagefraction{.001}

\shorttitle{Morpho-kinematic of galaxies at cosmic noon}

\shortauthors{Rigamonti F. et al.}  

\title[mode = title]{Morpho-kinematic of galaxies at cosmic noon}

\tnotetext[1]{} 


\author[1,2,3]{F. Rigamonti}[orcid=0000-0001-7880-8825]

\cormark[1]


\ead{fabio.rigamonti@inaf.it}

            
\author[1]{F. R. Ditrani}

\affiliation[1]{organization={INAF - Osservatorio Astronomico di Brera},
            addressline={via Brera, 28}, 
            city={Milano},
            postcode={20121}, 
            country={Italy}}
            
\affiliation[2]{organization={Como Lake centre for AstroPhysics (CLAP)},
            addressline={via Valleggio 11}, 
            city={Como},
            postcode={22100}, 
            country={Italy}}
        
\affiliation[3]{organization={INFN, Sezione di Milano-Bicocca},
            addressline={Piazza della Scienza 3}, 
            city={Milano},
            postcode={20126}, 
            country={Italy}}

\begin{abstract}
Recent studies of local galaxies highlight the need for high-resolution photometry and kinematics to accurately characterize galactic structures. From this, limitations emerged in the standard two-phase evolutionary scenario, where galaxies first form dispersion-supported bulges followed by secular disk growth. IFS surveys, combining photometry and kinematics, demonstrated that morphology and dynamics correlate weakly, with bulges and disks forming a continuum in their kinematic properties. This scenario is further supported by JWST observations. From analyses based on visual morphology emerged that mature galaxies with distinct morphological features exist at earlier times than expected. This, however, needs to be confirmed with high-resolution stellar kinematics observations.

We, therefore, propose to measure stellar kinematics at high spatial resolution for massive galaxies at cosmic noon to probe the build-up of central regions and to disentangle the true nature of structures inferred from visual morphology (e.g., disks and bulges with different degrees of rotation). This will enable studying how galaxy mass, kinematics, and star formation co-evolve, providing a complete census of galactic structures and their formation pathways. In this regard, the VESPER-SHARP instrument on the E-ELT is uniquely suited for this program, offering, within reasonable exposure times, rest-frame optical coverage at $z\sim2$, high spatial resolution, a suitable field of view, and multiplexing for simultaneous observations.

\end{abstract}





\maketitle



\section{Introduction: mass-budget measurements and their limitation}
\label{sec:introduction}

The Hubble tuning fork \citep[][]{Hubble_1926}{}{} traces multiple properties of local ($z \lesssim 0.1$) galaxies, including morphology (discs, bulges, bars), star formation (SF, traced by spiral arms), and kinematics (discs typically trace rotation) \citep[][]{Cappellari_2011,Kormendy_2012}{}{}. The correlation between these observables indicates that galaxies follow common evolutionary paths. This idea culminated in the two-phase model \citep[][]{Oser_2010}{}{}, where evolution is regulated by an early stage of gas collapse and mergers, followed by a secular phase with in-situ SF. The two phases of galaxy evolution are associated with the build-up of the most prominent stellar structures observed in the local galaxies: bulges and discs \citep[][]{Driver_2013}. In this context, bulges, formed via mergers, are considered tracers of the first violent phase of galaxy evolution, while discs indicate secular evolution and in-situ SF. 

This framework naturally motivated the widespread adoption of bulge–disc model as a standard tool to assess the structural constituents of galaxies identified via 2D imaging decomposition\footnote{In this work, we refer to bulges and disks identified via imaging decomposition as photometric bulges and photometric disks.} \citep[][]{Peng_2002,Gadotti_2009,Simard_2011,Lange_2016}{}{}. The underlying idea is to employ photometric decomposition on statistically complete samples from the widest photometric surveys \citep[e.g., GAMA, SDSS,][]{Driver_2009,SDSS} to measure the mass-budget of bulges and disks and, therefore, determine which is the dominant mechanism (i.e., mergers or in-situ SF) involved in galaxy evolution. Results coming from different, purely photometric, decomposition approaches broadly agree, equally splitting stellar mass into the bulge (or spheroid) and disc components \citep[][]{Moffet_2016,Thanjavur_2016,Bellstedt_2023}{}{}. From these studies emerged also that spheroidal galaxies (i.e., ideally diskless systems) dominated above $M_{\star}\simeq 10^{11}\rm \msun$, contributing up to 80\%-100\% of the total mass budget in that mass range. Even though these results may hint at a Universe in which mergers play a prominent role \citep[][]{van_Dokkum2010,Lopez_2012}{}{}, there may be biases.

The first issue arises from spatial resolution. Galaxies are rarely well described by bulge + disk systems. This has always been clear at large scales where the presence of non-axisymmetric structures—such as bars, rings, and spiral arms—complicates the picture. These components are often neglected under the assumption that they contribute only marginally to the total stellar mass and would, in most cases, be absorbed into the disc component without significantly affecting estimates of the secular mass budget. However, high-resolution imaging \citep[][]{Erwin_2015, Erwin_2021}, eventually combined with integral-field spectroscopy \citep[IFS;][]{Gadotti2020}, has revealed that low-resolution data can lead to substantial overestimates of bulge mass. Indeed, it has been shown that multiple central stellar structures, such as bars, inner discs, or nuclear components are blended together when observed in low resolution, artificially producing a compact, round central feature that is incorrectly interpreted as a photometric bulge. The latter, if associated with the stellar component forming as a consequence of violent processes, would lead to a misinterpretation of galaxies evolutionary picture.

The second limitation comes from the lack of correlation between morphology and kinematics. On scales of the galactic disk, IFS surveys have shown that a large fraction of visually classified ellipticals exhibit disk-like rotation \citep[][]{Emsellem2011, Cappellari_2016}. Similarly, photometrically identified bulges span a broad continuum of kinematic properties, ranging from dispersion-dominated (i.e., classical bulges) to rotation-dominated (i.e., pseudobulges or nuclear discs). Violent evolutionary processes are expected to leave a signature in stellar kinematics, increasing the contribution from random motion; therefore, not accounting for stellar dynamics while assessing the mass budget of galaxies would inevitably lead to biased results. Indeed, when galaxies are classified based on their global kinematic structure, the stellar mass residing in purely spheroidal systems amounts to less than 10\% of the total local mass budget \citep[][]{vandeSande2017,Guo_2020,Fraser-McKelvie_2022}{}{}. 

The picture depicted here comes from results on low-reshift galaxies, but to be further probed, it requires testing during the time in which galactic structure emergence is most prominent: the cosmic noon ($z\sim2$). Nowadays, this range of redshift can be probed only at moderate spatial resolution (e.g., James Webb Telescope, JWST) and, most importantly, without precise measurements of stellar kinematics. A major leap forward in this field will be provided by the next-generation near-infrared (NIR) IFSs such as VESPER-SHARP at the E-ELT. Such an instrument, by providing detailed measurements
of rest-frame optical spectra at  $z \gtrsim 2$ with an unprecedented spatial resolution, will allow morpho-kinematics analysis at cosmic noon, ultimately leading to a measurement of the mass-budget of galaxies and of their constituent through cosmic time.

\section{Morpho-kinematics at low redshift}
\label{sec:2}
When IFS observations are available, multiple approaches have been proposed to combine morphology and kinematics in order to obtain a more general and less biased decomposition of galactic components.  This has been done either by slicing datacubes into several narrow bands and applying bulge+disc decomposition on each of them \citep[][]{Johnston_2017,Mendez2019}{}{} or by assuming a photometric decomposition and extracting the bulge and disc spectra directly from the datacubes \citep[][]{Oh_2016,Tabor_2017,Tabor_2019,Pak_2021}. These methods focus on extracting the stellar population and the star formation histories of bulges and discs, but always assume decomposition purely based on photometry (i.e., without considering galactic kinematics).

On the opposite side, alternative approaches focus on modeling galaxy surface brightness and its line-of-sight kinematics as the superposition of purposely weighted orbital families \citep[][]{Schwarzschild1979}. This obits-based method builds a full dynamical model of the system, measuring the visible and dark mass distribution of galaxies and the properties (e.g., anisotropy, circularity) of their orbital structure, but neglects any decomposition based on visible stellar structures without allowing proper recovery of bulge and disc properties.   

\begin{figure}
    \centering
    \includegraphics[scale=0.6]{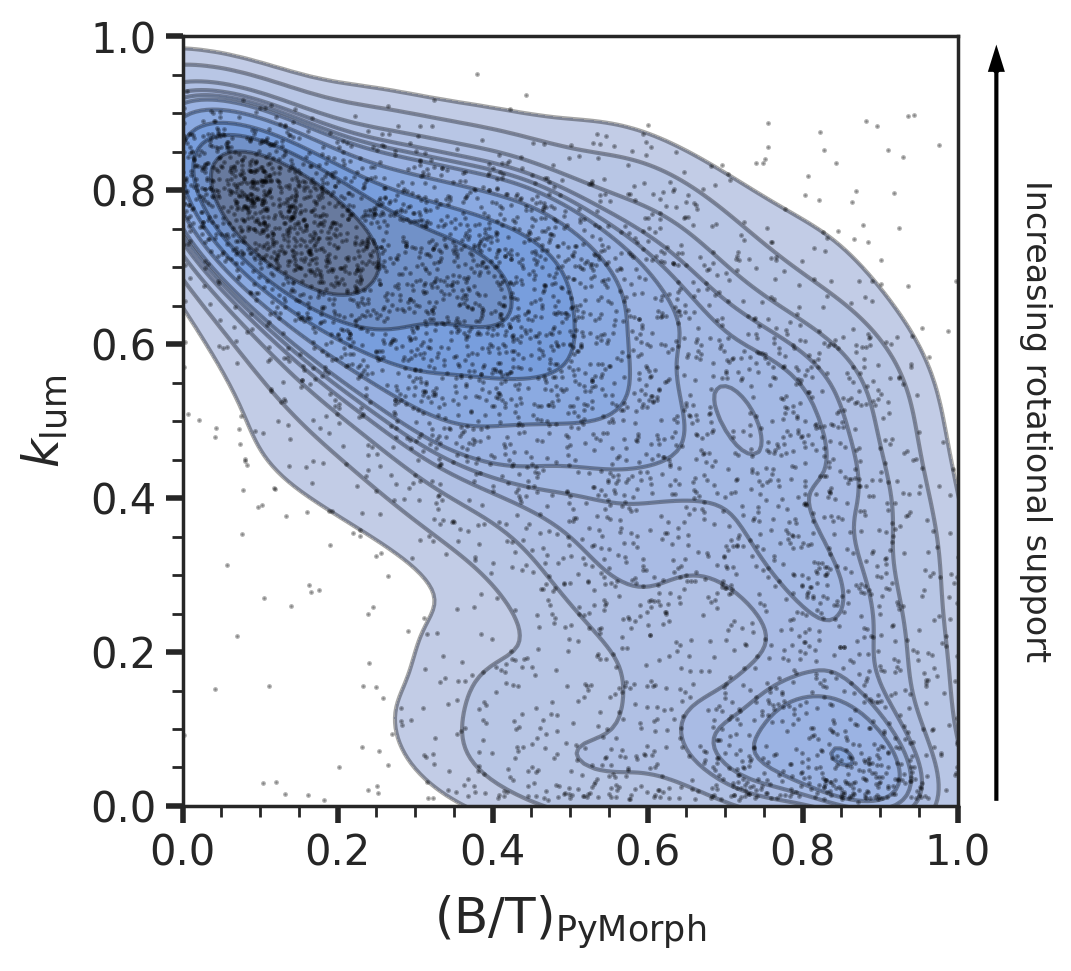}
    \caption{Comparison between the luminosity-weighted kinematic tracer $k_{\rm{lum}}$ \citep[vertical axis, see ][]{Rigamonti2024} and a purely photometric bulge-to-total ratio taken from \cite{Sanchez2022} (horizontal axis), on the MaNGA sample analyzed in \citet[][]{Rigamonti2024}. The contours (grey lines filled with the blue shaded colours) are drawn at probability levels with a constant spacing of 0.1, while the black dots represent the data. The vertical arrow on the right side qualitatively explains the meaning of the $k_{\rm{lum}}$ parameters: a greater $k_{\rm{lum}}$ value corresponds to higher rotational support within the galaxy.}
    \label{fig:klum_vs_BT}
\end{figure}

In \cite{Rigamonti_2022}, we presented \textsc{bang}\footnote{\url{https://pypi.org/project/BANGal/}}, a code purposely designed for a simultaneous and self-consistent bulge+disc decomposition of galaxy photometry and kinematics. Being informed by the kinematics \textsc{bang}'s decomposition suffers minimally from the possible biases of purely photometric decomposition.
\textsc{bang} bases its parameter estimation on a robust Bayesian algorithm \citep[i.e., nested sampling][]{Skilling2004}{}{} combined with a GPU parallelisation strategy for high computational performance, representing an optimal tool for automated analysis on large galaxy samples. This allowed us to present the first morpho-kinematic bulge+disc decomposition obtained through galaxy dynamical models \citep[][]{Rigamonti_2023}{}{} of the largest, to date, IFS survey \citep[i.e. the Mapping Nearby Galaxies at Apache Point Observatory, MaNGA, survey][]{Bundy_2015}{}{}. The analysis has been used in \citet[][]{Rigamonti2024} to precisely quantify the mass-budget of galaxies as a function of global galaxy kinematics, individual components kinematics, SF, and mass. In that work, it was demonstrated that the distinction between dispersion-supported bulges and rotationally-supported disks is not sharp. Although galaxies may appear to consist of these two discrete classes, a relevant fraction exhibits intermediate behavior, such as central regions with significant rotation or extended disks that retain a non-negligible level of dispersion support. This is partially summarized by Fig.~\ref{fig:klum_vs_BT}, where a photometrically determined bulge-to-total light ratio \citep[B/T, ][]{Dominguez2020} has been compared with a measurement of the intrinsic galaxy kinematics \citep[$k_{\rm lum}$, ][]{Rigamonti2024}. Although a correlation between the two reported quantities exists, the scatter is significant. This implies that morphology traces kinematics only to a first order, as galaxies with similar B/T ratios can almost span the full range of dynamical states. When SF is added to the picture (see Fig.~7 of \citealt{Rigamonti2024}), the mismatch becomes even more evident. For galaxies with $10^{9.5} \lesssim ~ \msun \lesssim 10^{11}$, SF shows little correlation with the true, mass-weighted kinematic structure. This suggests that structural transformation\footnote{We define structural transformation as any process that converts a rotationally supported system into a dispersion-supported one.} is not directly tied to the quenching of SF, effectively ruling out morphological quenching as a major driver in galaxy evolution. 

From an evolutionary perspective, these findings imply that the classical two-phase model captures only the macroscopic trend in galaxy evolution. To account for the observed diversity in stellar kinematics, multiple and possibly concurrent evolutionary pathways must be considered. Overall, this highlights the importance of a morpho-kinematic approach, which passes through dynamical modeling, for probing the internal structure of galaxies and their evolutionary histories.

\section{Morpho-kinematics at high redshift with VESPER-SHARP}
The results presented in Sec .~\ref {sec:2} are limited to the local Universe (i.e., $z\lesssim0.1$) and therefore, allow inference on the evolution of galaxies a posteriori (i.e., after the evolutionary process has already taken place). A natural next step is to extend such analyses to higher redshifts while maintaining both the spatial resolution and the signal-to-noise required to measure stellar kinematics on sub-kpc scales. Achieving this would allow the approach described in Sec.~\ref{sec:2} to be applied directly to galaxies in earlier cosmic epochs. Ultimately, this is the goal of our science case: to enable accurate morpho-kinematic modeling of distant galaxies and obtain direct constraints on the physical mechanisms driving the formation and assembly of galactic structures as they occur (i.e., at cosmic noon).

Moreover, obtaining the measurement mentioned above is already crucial, and it will become a necessity in the following years. Indeed, the most recent discoveries from the James Webb Space Telescope (JWST) suggest that even at relatively early times (i.e., $z\simeq3$), morphologically mature galaxies are common. As a matter of fact, multiple works indicate that the Hubble Sequence was already in place as early as one billion years after the Big Bang, with disk galaxies being the dominant class above $10^9~\msun$ up to $z=6$ \citep[][]{Ferreira2023,Kartaltepe2023,Huertas-Company2024}. 

Notably, these results mostly come from analysis based on visual morphology and do not consider spatially resolved information from stellar kinematics. As detailed in Sec.~\ref{sec:introduction}, neglecting this information can substantially bias our understanding of the true internal structure of (JWST) galaxies, ultimately leading to misleading conclusions about their evolutionary pathways. A few studies have already attempted to infer the intrinsic shapes and structural properties of JWST galaxies using purpose-trained probabilistic models \citep[][]{Pandya2024}, combining imaging with ionized-gas kinematics \citep[][]{Rizzo2021}, or extracting stellar kinematics from slit spectroscopy \citep[][]{Slob2025}. In all these cases, further assumptions (e.g., axis-symmetry) are made to extend what is observed into a conclusion about intrinsic stellar shapes or velocity fields. Just in one case, spatially resolved stellar kinematics have been observed with JWST/NIRSpec IFS \citep[][]{D'Eugenio2024}, demonstrating the feasibility of such observation with JWST, although at relatively low spatial resolution. Indeed, JWST/NIRSpec IFS will provide, in the optimal scenario (i.e., $0.1''$ arcsec/pixel), about 25 pixels within the effective radius of a typical $z=2$ galaxy (i.e., $R_e=0.250''\simeq 2~{\rm kpc}$). Such resolution is likely sufficient for recovering global kinematic quantities (e.g., the luminosity-weighted spin parameter $\lambda_{R,e}$), but too low to identify and characterize the kinematic features of the observed stellar structures.

\begin{figure*}
    \centering
    \includegraphics[scale=0.6]{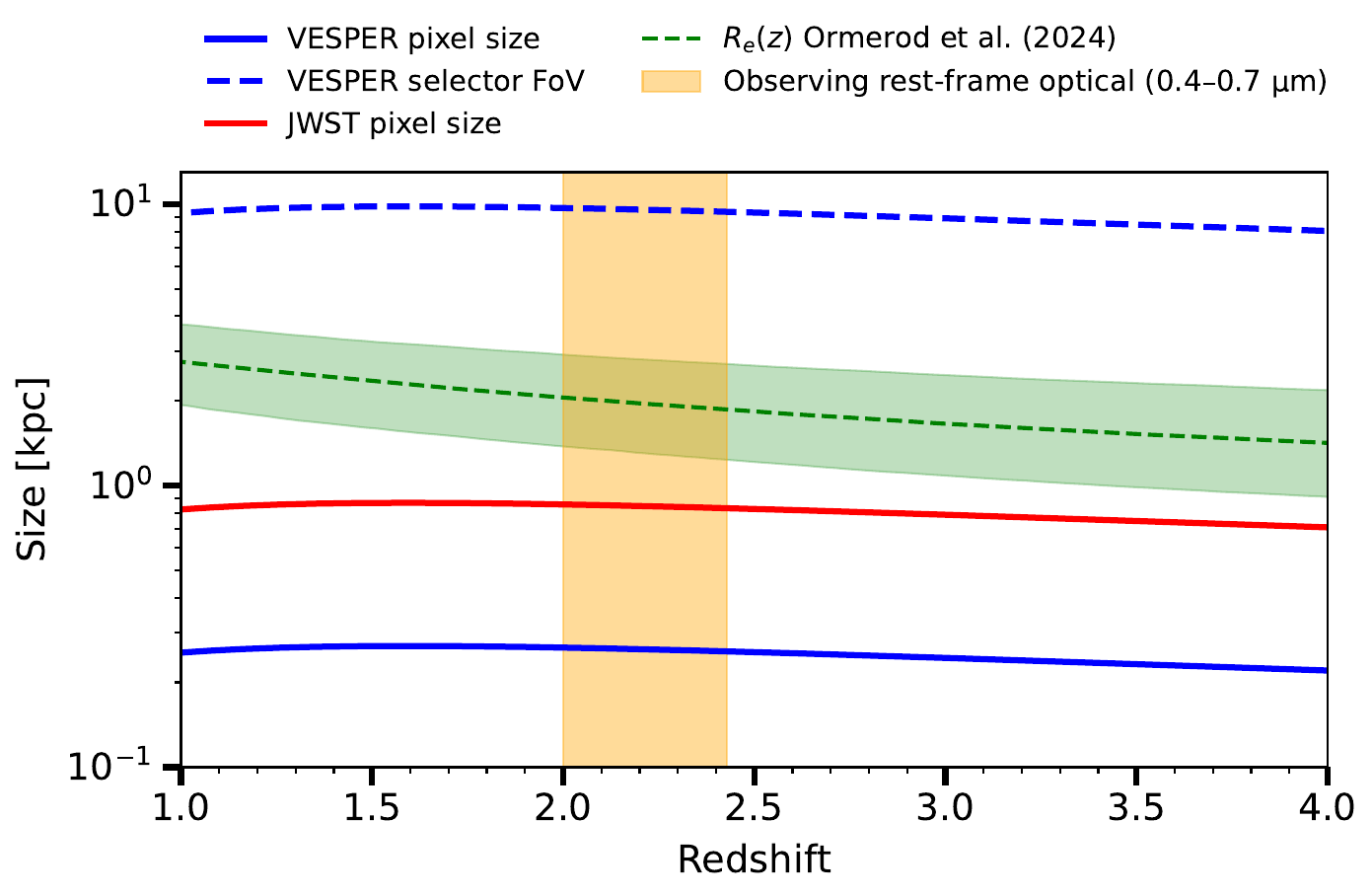}
    \caption{Relevant sizes as a function of redshift. Solid blue and red lines are the pixel sizes of VESPER and JWST IFU, respectively. The dashed blue line is the FoV of a VESPER Integral Field Selectors (IFSs). The green dashed line and the associated shaded area represent the evolution of galaxies' effective radius with redshift taken from \cite{Ormerod2024}. The orange shaded area represents the redshift range within which the whole rest frame optical ($0.4\mu m-0.7\mu m$) is included in the observed window of VESPER.}
    \label{fig:resolution vs redshift}
\end{figure*}

These limitations will be solved with the advent of next-generation near-infrared (NIR) IFSs, such as VESPER-SHARP at the E-ELT, which will allow detailed measurements of rest-frame optical spectra at $z\gtrsim2$ with an unprecedented spatial resolution.

As summarized in Fig.~\ref{fig:resolution vs redshift}, VESPER will be particularly suited for fulfilling this task for multiple reasons:
\begin{itemize}
    \item \textbf{Unprecedented spatial resolution}\\
    Fig.~\ref{fig:resolution vs redshift} reports the pixel scale as a function of redshift for VESPER (blue solid line) in comparison with JWST-NIRspec (red solid line) and with the typical effective radius of a galaxy \citep[green dashed line, ][]{Ormerod2024}. From the comparison of these 3 lines, it is clear that VESPER will have enough spatial resolution to characterize galactic central regions at $z \gtrsim 2$, obtaining more than 100 different spectra within $1R_e$ for an individual target.
    \item \textbf{Observed wavelenght range}\\
     Our purpose is to measure stellar kinematics for galaxies at cosmic noon. Measuring stellar kinematics of galaxies is typically achieved via full-spectral fitting \citep[e.g., pPXF][]{Cappellari_2016} of spectra to extract moments of the line-of-sight velocity distribution. This approach works best in the rest-frame optical range, where the emission is dominated by stars showing prominent absorption features whose signature is used to precisely recover stellar kinematics. In particular, we report with the orange shaded area in Fig.~\ref{fig:resolution vs redshift} the redshift range within which the whole rest-frame optical ($0.4-0.7~{\rm \mu m}$) falls in the observed VESPER wavelength coverage (i.e., $1.2-2.4~{\rm \mu m}$). The optimal redshift is in between $2\lesssim z \lesssim 2.5$, overlapping with the cosmic noon, where galaxies evolve into their mature form, and optimally aligning with the science case presented in this work.
    \item \textbf{Flexible Field-of-View}\\
    VESPER is a multi-IFU composed of 12 probes called Integral Field Selectors. Each selector has a FoV of $\sim 1.7'' \times 1.5'' $ that can be placed in a Cartesian grid to provide a full field of $\sim 24'' \times 70'' $. The size of a single selector (converted into spherical radius) is reported with a blue dashed line in Fig.~\ref{fig:resolution vs redshift}, demonstrating that, already with a single selector, it will be possible to cover more than $1~R_e$. Still, the VESPER multiplexing capability will be useful. For extended galaxies, placing selectors one beside the other will allow measurements of the entire galaxy with a single observation. Also, by precisely disposing the selectors in the full FoV it will be possible to target multiple galaxies simultaneously. This will be particularly relevant in crowded regions (e.g., clusters and proto-clusters). For instance, by considering galaxies with stellar masses above $M_* > 10^{10.5} M_\odot$ in the redshift range $1.8 < z < 2.5$ for two representative clusters at cosmic noon \citep[JKCS 041, XLSSC 122,][]{newman2014,willis2020} we estimate a number density of $7-10$ galaxies with H-band magnitudes brighter than $21.5$ allowing us to make full use of VESPER multiplexing capabilities.   
\end{itemize}

In Tab.~\ref{tab:parameters} we report the feasibility of our proposed observation. Our primary targets are bright galaxies (Mag in H-band $\sim 21$) at $z\sim2$ to be observed with VESPER to measure stellar kinematics out to $1~R_e$. To evaluate the feasibility, we employed the dedicated exposure time calculator (Version 0.6), computing the exposure time needed to reach \textcolor{black}{S/N per spatial pixel} $\sim 10$ at $1~R_e$. This has been done in the range of magnitudes between 20.5-21.5 for disk-like ($n=1$) and bulge-like ($n=4)$ systems. In all cases, we adopted a spectral binning of 4 pixels. In the table, we also report the S/N reached in the brightest pixels. Observations for the faintest sources are demanding, but larger spectral or spatial binning (e.g., Voronoi bins) can be considered for mitigating the requested exposure times or for measuring stellar kinematics up to radii larger than $1~R_e$. Taking these considerations into account, we conclude that the observations required for the proposed science case are realistically achievable with VESPER.

\begin{table}
	\caption{Table simulating VESPER exposure time for reaching a S/N \textcolor{black}{per spatial pixel} of 10 at approximately $1 R_e$. All simulations are done at $z=2$ and consider a Sersic brightness profile with a $R_e = 0.25''$ ($\simeq 2$ kpc at $z=2$). The tables report (from left to right) the assumed H-band magnitude, the Sersic index n, the maximum S/N reached, and the required exposure time.}
	\label{tab:parameters}
	\small{\begin{tabular}{cccc} 
        \hline\vspace{-0.75em}\\
		Mag H-band & n  & Max S/N & Exposure Time [hrs]\vspace{0.2em}\\
        \hline\vspace{-0.75em}\\
        20.5 & 1 or 4 & 21 or 150 & 5 or 14  \\
        21.0 & 1 or 4 & 28 or 160 & 17 or 30 \\
        21.5 & 1 or 4 & 31 or 170 & 47 or 67\\  \hline
	\end{tabular}}
\end{table}
    
\section{Summary and conclusions}
Studies of galaxies in the local Universe have highlighted the need for high spatial resolution kinematics to probe the true nature of galactic structures \citep[][]{Gadotti2020}. Moreover, contrary to earlier expectations, observations from IFS surveys \citep[][]{Bundy_2015} combining photometry and kinematics have shown that the correlation between morphology and kinematics is significantly weaker than previously thought \citep[][]{Rigamonti2024}. These results, once put in a coherent framework incorporating star formation and stellar populations, reveal a Universe in which the standard two-phase evolutionary scenario \citep[][]{Oser_2010,Driver_2013} is challenged. This became even more evident with recent JWST observations, which indicate a higher fraction of visually identified disks at early cosmic times. Accurately confirming the intrinsic morphology of the structures observed in these systems is therefore critical to build a reliable model of galaxy evolution.

With this science case, mimicking what has been done in the local Universe, we proposed a set of observations to measure stellar kinematics at high spatial resolution for massive galaxies at cosmic noon. High spatial resolution is needed to probe the build-up of galactic central regions, while spatially resolved kinematics is crucial to understand the true nature of structures inferred from visual inspection (e.g., disks or bulges with more or less rotation). In practice, once this is done on a relevant sample of galaxies, and consistently in a relatively broad range of redshifts, it will allow an assessment of how the mass budget of galaxies depends on kinematics, star formation, and redshift, building a complete census of galactic structures and how they form and evolve. 

Thanks to the combination of its unique capabilities, VESPER-SHARP on the E-ELT is the ideal instrument to address this scientific goal. Its wavelength coverage matches the rest-frame optical at $z\sim2$, its high spatial resolution, and the FoV of each Integral Field Selector are well matched to the scales required to resolve galaxy central regions while still capturing the full extent of a system beyond $1~R_e$. In addition, VESPER multiplexing capability provides substantial flexibility in observational strategy, enabling both the study of the outer regions of extended galaxies and the simultaneous acquisition of spatially resolved data for multiple galaxies in clusters. Our science case also benefits from a natural synergy with any observational program that exploits VESPER IFS capabilities to observe galaxies: any such dataset includes, by construction, the information needed for stellar kinematic analysis, and thus can directly contribute to this project. Finally, preliminary exposure-time estimates indicate that the proposed observations are feasible for typical bright galaxies at $z\sim2$, providing strong support for the feasibility and scientific impact of the proposed project.


\bibliographystyle{aa}
\bibliography{example}






\end{document}